\documentclass{iopconfser}
\usepackage{enumerate}
\usepackage{graphicx}
\usepackage{fancyhdr}
\begin{document}
\title{Analytical Kink-Type Solutions and Streak Formation in Turbulent Channel Flow}
\author{Alex Fedoseyev}
\affil{Ultra Quantum Inc., Huntsville, Alabama, USA}
\email{af@ultraquantum.com}

\begin{abstract}

An analytical framework for turbulent channel flow is developed based on the Alexeev hydrodynamic equations, focusing on the coupled behavior of streamwise and transverse velocity components. The mean streamwise velocity is represented as a superposition of a laminar (parabolic) component and a nonlinear turbulent contribution, yielding velocity profiles that agree with experimental data from channel and pipe flows over a wide range of Reynolds numbers, $3\times10^3 \le Re \le 3.5\times10^7$, with deviations of approximately $1\%$ at moderate Reynolds numbers and up to $3\%$ at the highest Reynolds numbers.

The transverse velocity component is analyzed using a simplified form of the governing equations, leading to analytical expressions that capture its dominant spatial structure. The coupling between transverse velocity and streamwise momentum is then examined, revealing that the streamwise turbulent component admits a family of kink-type solutions.

These solutions exhibit localized monotonic transitions separating regions of nearly uniform velocity and are interpreted as analytical representations of streamwise streaks. The model predicts characteristic streak properties, including spacing, thickness, intensity, and streamwise extent, which are shown to be consistent in order of magnitude with experimental observations of near-wall streaks.

The results provide a unified analytical description of mean velocity profiles, secondary flows, and streak formation in wall-bounded turbulence, and suggest a mechanism linking transverse velocity fluctuations to the emergence of coherent streamwise structures.


\end{abstract}
\section{Introduction}

Turbulent wall-bounded flows, such as channel and pipe flows, exhibit a rich multiscale structure characterized by strong anisotropy, near-wall coherent structures, and the presence of secondary motions. Among the most prominent features are streamwise streaks and quasi-streamwise vortices, which play a central role in momentum transport and in the self-sustaining cycle of near-wall turbulence. Despite decades of experimental, numerical, and theoretical work, a predictive analytical description that captures both the mean flow and the formation of these structures remains an open challenge.

Classical approaches to wall-bounded turbulence rely on empirical closures, asymptotic scaling arguments, or direct numerical simulations. While these methods have achieved considerable success in describing statistical properties of turbulence, they do not generally yield closed-form analytical solutions capable of simultaneously representing the mean velocity profile and coherent structures such as streaks. In particular, the coupling between streamwise flow, transverse motion, and secondary structures is typically described only indirectly through Reynolds stresses or modeled transport equations.

In recent work, an approximate analytical solution for turbulent channel flow was developed using the Alexeev hydrodynamic equations \cite{Fedoseyev_2023,Fedoseyev_2025}. In this framework, the mean streamwise velocity is represented as a superposition of a parabolic (laminar) component and a rapidly varying nonlinear contribution. This representation was shown to reproduce experimental mean velocity profiles from canonical pipe and channel flows over a wide range of Reynolds numbers, $3\times10^3 \le Re \le 3.5\times10^7$, with quantitative errors of approximately $1\%$ for $3\times10^3 \le Re \le 4.5\times10^4$ and about $3\%$ at $Re \approx 3.5\times10^7$.

Subsequently, governing equations for the transverse velocity component were derived and approximate analytical solutions were obtained \cite{Fedoseyev_2024}. These results suggest that the Alexeev hydrodynamic framework can consistently describe not only the mean streamwise flow but also transverse motions associated with secondary flow structures.

The objective of the present work is to extend this analytical framework by examining the coupled behavior of the streamwise and transverse velocity fields. In particular, we show that the streamwise velocity admits a family of kink-type solutions, the monotonic transition solutions connecting two distinct asymptotic states, characterized by a localized region of large gradient separating nearly uniform regions of the flow, whose properties depend on the amplitude and structure of the transverse velocity field. These solutions are interpreted as streamwise streaks propagating with different velocities, providing a direct analytical link between transverse motion and streak formation.

The predicted characteristics of these structures are compared with available experimental observations of near-wall streaks, including their spatial organization and propagation behavior. In this sense, the present study provides, to the authors' knowledge, the first analytical description within this framework of the interaction between secondary flow and streak formation in turbulent channel flow.

The results suggest a consistent analytical picture in which secondary motions modulate the structure and dynamics of streamwise streaks, offering insight into possible mechanisms underlying turbulence generation in wall-bounded flows.

The paper is organized as follows. Section 2 summarizes the governing equations and the previously derived analytical solution for the streamwise velocity. Section 3 presents the equations and solutions for the transverse velocity component. Section 4 analyzes the coupled system and derives the family of kink-type solutions. Section 5 compares the theoretical predictions with experimental observations and numerical results. Conclusions are given in Section 6.
%
%

\section{Analytical solution for streamwise velocity \label{sec:govan}}

%
%
\subsection{Governing equations}
\label{sec:AHE}

The Alexeev hydrodynamic equations (AHE) \cite{Alexeev_1994, Alexeev_2004} are written in nondimensional form as
%
%
\begin{equation}
\frac{\partial \mathbf{V}}{\partial t}
+ (\mathbf{V}\cdot\nabla)\mathbf{V}
- \frac{1}{Re}\nabla^2 \mathbf{V}
+ \nabla p
- \mathbf{F}
=
\tau \left[
\frac{\partial^2 \mathbf{V}}{\partial t^2}
+ 2 \frac{\partial}{\partial t}(\nabla p)
+ \nabla^2 (p \mathbf{V})
+ \nabla (\nabla \cdot (p \mathbf{V}))
\right],
\label{eq:mom}
\end{equation}
with the continuity equation
\begin{equation}
\nabla \cdot \mathbf{V}
=
\tau \left[
2 \frac{\partial}{\partial t}(\nabla \cdot \mathbf{V})
+ \nabla \cdot \big((\mathbf{V}\cdot\nabla)\mathbf{V}\big)
+ \nabla^2 p
- \nabla \cdot \mathbf{F}
\right].
\label{eq:cont}
\end{equation}

Here $\mathbf{V}$ and $p$ are the nondimensional velocity and pressure, respectively, $Re=U_0 L_0/\nu$ is the Reynolds number, and $\mathbf{F}$ is a body force. The parameter $\tau$ introduces an additional characteristic scale and is defined as
\begin{equation}
\tau = \delta^2 Re,
\label{eq:tau}
\end{equation}
where
\begin{equation}
\delta = \frac{\sqrt{\tau_0 \nu}}{L_0}.
\label{eq:delta}
\end{equation}

The AHE reduce to the incompressible Navier-Stokes equations in the limit $\tau \to 0$. The parameter $\delta$ introduces an additional length scale that is treated here as a fluid property.

For solid walls, the pressure boundary condition is
\begin{equation}
(\nabla p - \mathbf{F})\cdot \mathbf{n} = 0,
\end{equation}
where $\mathbf{n}$ is the wall-normal vector.

%
%
\subsection{Analytical solution for streamwise velocity}
\label{sec:ansolu}

We consider steady two-dimensional channel flow and assume that all quantities depend only on the wall-normal coordinate $y$. The streamwise pressure gradient $dp/dx$ is taken to be constant.

%
%
\subsubsection{Simplified AHE for 2D stationary channel flow}
\label{sec:2d_ahe}

Neglecting time derivatives and retaining only the fluctuation term $\tau {d^2 p  /dy^2}$, the governing equations reduce to
\begin{equation}
V \frac{dU}{dy} - \frac{1}{Re}\frac{d^2 U}{dy^2} + \frac{dp}{dx} = 0,
\label{eq:2d_mom_U}
\end{equation}
\begin{equation}
V \frac{dV}{dy} - \frac{1}{Re}\frac{d^2 V}{dy^2} + \frac{dp}{dy} = 0,
\label{eq:2d_mom_V}
\end{equation}
\begin{equation}
\frac{dV}{dy} = \tau \frac{d^2 p}{dy^2}.
\label{eq:2d_cont_ahe}
\end{equation}

To construct an analytical solution, we represent the velocity field as a superposition of laminar and turbulent components:
\begin{equation}
U(y) = U_0\left[\gamma U^T(y) + (1-\gamma) U^L(y)\right],
\label{eq:2d_usol}
\end{equation}
\begin{equation}
V(y) = \gamma V^T(y),
\label{eq:2d_vsol}
\end{equation}
where $\gamma$ is a weighting parameter.

%
%
\subsubsection{Laminar component}

The laminar component satisfies
\begin{equation}
- \frac{1}{Re} \frac{d^2 U^L}{dy^2} + \frac{dp}{dx} = 0,
\end{equation}
with boundary conditions $U^L(0)=0$ and $U^L(1/2)=U_0$. The solution is
\begin{equation}
U^L(y) = U_0 \, 4y(1-y),
\end{equation}
and $V^L=0$, which corresponds to the classical parabolic profile.

%
%
\subsubsection{Turbulent component}

The turbulent component satisfies
\begin{equation}
V^T \left(\frac{dU^T}{dy} + \alpha \frac{dU^L}{dy}\right)
- \frac{1}{Re}\frac{d^2 U^T}{dy^2} = 0,
\label{eq:2d_mom_UT}
\end{equation}
\begin{equation} \label{eq:2d_mom_VT}
V^T \frac{dV^T}{dy} - \frac{1}{Re}\frac{d^2 V^T}{dy^2} + \frac{dp^T}{dy} = 0,
\end{equation}
\begin{equation} \label{eq:2d_cont_VT}
\frac{dV^T}{dy} = \delta^2 Re \frac{d^2 p^T}{dy^2},
\end{equation}
where $\alpha = (1-\gamma)/\gamma$.

Following Ref.~\cite{Fedoseyev_2023}, an approximate solution for $V^T$, was obtained from equation (\ref{eq:2d_mom_VT}) by dropping the nonlinear term $V^T V^T_y$, differentiating it with respect to $y$, and substituting the expression for $ p^T_{yy}$ into  equation (\ref{eq:2d_cont_VT}), which leads to
\begin{equation}
\frac{d^3 V^T}{dy^3} - \frac{1}{\delta^2}\frac{dV^T}{dy} = 0.
\end{equation}

A representative solution  is
\begin{equation}
V^T = \frac{1}{\delta Re}\left(1 - e^{y/\delta}\right).
\label{eq:2d_VT_sol1}
\end{equation}

Substituting this expression into Eq.~(\ref{eq:2d_mom_UT}) and neglecting the small term containing $\alpha dU^L/dy$ yields the approximate solution
\begin{equation}
U^T = U_0\left(1 - e^{1 - e^{y/\delta}}\right).
\label{eq:UTsol}
\end{equation}

%
%
\subsection{General  solution for streamwise velocity}
The total velocity profile is then given by
\begin{equation}
U(y) = U_0 \left[
\gamma\left(1 - e^{1 - e^{y/\delta}}\right)
+ (1-\gamma) 4y(1-y)
\right].
\label{eq:AHEsol}
\end{equation}

The determination of the parameter $\gamma$  through the principle of minimal total viscous dissipation was proposed in \cite{Fedoseyev_2024b}. The value of $\gamma$ obtained in this way provided the best fit to the experimental data with typical values $\gamma$ between  0.6 and 0.7 for the flows considered.

%
%
\subsection{Improved solution for streamwise velocity}

The second solution for $V^T$, when the sign of $\delta=-\sqrt{\delta^2}$ is taken as negative, was considered in \cite{Fedoseyev_2025}:

\begin{equation}
\label{eq:2d_VT_sol2}
V^T= -\frac{1}{\delta Re}\left({1-e^{-y/\delta}}\right).
\end{equation}

The linear combination of $U^T$ solutions, corresponding to the $V^T$ solutions (\ref{eq:2d_VT_sol1}), (\ref{eq:2d_VT_sol2}), gives more accurate results for $U$ in over wide range of Reynolds number from $3\times10^3$ to $3.5\times10^7$,  comparing well with experimental data from multiple sources \cite{Doorne_2007, Nikuradse_1932, Pasch_2024, Wei_1989, Zagarola_1996}, with quantitative errors of approximately $1\%$ for $3\times10^3 \le Re \le 4.5\times10^4$ and about $3\%$ at $Re \approx 3.5\times10^7$ \cite{Fedoseyev_2025}, Figure 1.  

%
%

\section{Governing equations and transient solutions for the transverse velocity}
\label{sec:transition}

We consider two-dimensional incompressible flow and derive equations governing the transverse velocity component. The AHE system (\ref{eq:mom})--(\ref{eq:cont}) is simplified by neglecting higher-order nonlinear fluctuation terms, retaining only the leading contributions.
%
%

\subsection{Governing equations for 2D flow}
\label{sec:AHE2D}

The resulting equations are
\begin{equation}
u_x + v_y = \tau \left[ 2(u_{xt} + v_{yt}) + \nabla^2 p \right],
\label{eq:cont0}
\end{equation}
\begin{equation}
u_t + u u_x + v u_y + p_x = \frac{1}{Re} \nabla^2 u + \tau (u_{tt} + 2 p_{xt}),
\label{eq:u_mom}
\end{equation}
\begin{equation}
v_t + u v_x + v v_y + p_y = \frac{1}{Re} \nabla^2 v + \tau (v_{tt} + 2 p_{yt}).
\label{eq:v_mom}
\end{equation}

%
%
\subsection{Reduction for channel flow}
\label{sec:trans2D}

For channel flow we assume $u_x = v_x = 0$ and $p_x=\mathrm{const}$, and drop the fluctuation term in $u$-momentum equation, assuming is it small. The equations reduce to
\begin{equation}
v_y = \tau (2 v_{yt} + p_{yy}),
\label{eq:contch}
\end{equation}
\begin{equation}
u_t + v u_y + p_x = \frac{1}{Re} u_{yy},
\label{eq:u_momch}
\end{equation}
\begin{equation}
v_t + v v_y + p_y = \frac{1}{Re} v_{yy} + \tau (v_{tt} + 2 p_{yt}).
\label{eq:v_momch}
\end{equation}

Integrating Eq.~(\ref{eq:contch}) with respect to $y$ gives
\begin{equation}
p_y = \frac{1}{\tau} v - 2 v_t,
\label{eq:p4a}
\end{equation}
where the integration constant vanishes due to the wall conditions.

Differentiating in time yields
\begin{equation}
p_{yt} = \frac{1}{\tau} v_t - 2 v_{tt}.
\label{eq:p5}
\end{equation}

Substituting Eqs.~(\ref{eq:p4a}) and (\ref{eq:p5}) into Eq.~(\ref{eq:v_momch}) results in
\begin{equation}
v_{tt} - \frac{1}{\tau} v_t =
\frac{1}{3 \tau Re} v_{yy}
- \frac{1}{3 \tau} v v_y
- \frac{1}{3 \tau^2} v.
\label{eq:mom2d0}
\end{equation}

%
%
\subsection{Linearized transient equation}

Neglecting the nonlinear term $v v_y$ (valid when the transverse velocity is small compared to the streamwise component) yields
\begin{equation}
v_{tt} - \frac{1}{\tau} v_t =
\frac{1}{3 \tau Re} v_{yy}
- \frac{1}{3 \tau^2} v.
\label{eq:mom2d}
\end{equation}

%
%
\subsection{Separation of variables}

We seek solutions of the form
\begin{equation}
v(y,t) = g(y)\,h(t).
\end{equation}

Substitution into Eq.~(\ref{eq:mom2d}) leads to
\begin{equation}
\frac{h_{tt} - \frac{1}{\tau} h_t + \frac{1}{3\tau^2} h}{h}
=
\frac{1}{3\tau Re} \frac{g_{yy}}{g}
= -C,
\end{equation}
where $C>0$ is a separation constant.

This yields
\begin{equation}
h_{tt} - \frac{1}{\tau} h_t + \left(C + \frac{1}{3\tau^2}\right) h = 0,
\label{eq:h3}
\end{equation}
\begin{equation}
g_{yy} + 3\tau Re\, C\, g = 0.
\label{eq:g3}
\end{equation}

%
%
\subsection{Spatial solution}

Imposing no-penetration boundary conditions $v=0$ at the walls gives
\begin{equation}
g(y) = \sum_{n=1}^{\infty} g_n \sin\left(\frac{n\pi y}{L}\right),
\end{equation}
with
\begin{equation}
C = \frac{n^2 \pi^2}{3 L^2 \tau Re}.
\end{equation}

%
%
\subsection{Temporal solution}

Equation (\ref{eq:h3}) is a linear second-order ODE with constant coefficients. For the present parameters, the discriminant is negative, and the solution takes the form
\begin{equation}
h(t) = e^{t/(2\tau)} \left[ A_n \sin\left(\frac{\lambda_n t}{2}\right)
+ B_n \cos\left(\frac{\lambda_n t}{2}\right) \right],
\end{equation}
where
\begin{equation}
\lambda_n =
\frac{1}{\delta Re \sqrt{3}}
\sqrt{\frac{n^2 \pi^2}{L^2} + \frac{1}{\delta^2}}.
\end{equation}

%
%
\subsection{General solution}

The transverse velocity is therefore
\begin{equation}
v(y,t) =
e^{t/(2\delta^2 Re)}
\sum_{n=1}^{\infty}
\left[
A_n \sin\left(\frac{\lambda_n t}{2}\right)
+ B_n \cos\left(\frac{\lambda_n t}{2}\right)
\right]
\sin\left(\frac{n\pi y}{L}\right).
\label{eq:gensol_v}
\end{equation}

The coefficients are determined from initial conditions.

The solution describes oscillatory behavior modulated by an exponential factor. The growth term reflects the linearized nature of the model; in practice, nonlinear effects are expected to limit the amplitude of transverse velocity fluctuations.
This behavior is consistent with transient amplification mechanisms observed in wall-bounded turbulence.
%
\section{Analysis of the coupled system and kink-type solutions}
\label{sec:kink}
\subsection{Quasi-steady approximation for the streamwise component}

The governing equation for the turbulent streamwise velocity component can be written as
\begin{equation}
\frac{\partial U^T}{\partial t}
+ V^T(y,t)\frac{\partial U^T}{\partial y}
- \frac{1}{Re}\frac{\partial^2 U^T}{\partial y^2} = 0.
\end{equation}

The transverse velocity is assumed to have the form
\begin{equation}
V^T(y,t) = A \sin\left(\frac{\lambda_n t}{2}\right)\sin(n\pi y),
\end{equation}
where $\lambda_n$ defines the characteristic temporal frequency.

Two characteristic timescales can be identified. The diffusive timescale associated with the streamwise velocity is
\begin{equation}
T_0 \sim \frac{L_0^2}{\nu},
\end{equation}
while the transverse velocity varies on the timescale
\begin{equation}
T_V \sim \frac{4\pi}{\lambda_n}.
\end{equation}

For the parameters considered here, $T_V \ll T_0$, indicating that the transverse velocity varies rapidly compared to the evolution of $U^T$. In this regime, the streamwise velocity responds quasi-steadily to the instantaneous transverse forcing.

Consequently, to leading order, the time derivative $\partial U^T/\partial t$ can be neglected, yielding the reduced equation
\begin{equation} \label{eq:UTy}
V^T(y,t)\frac{d U^T}{d y}
- \frac{1}{Re}\frac{d^2 U^T}{d y^2} = 0.
\end{equation}

This equation describes a quasi-steady balance between transverse advection and viscous diffusion, with the rapid temporal oscillations of $V^T$ entering only through its effective spatial structure.
This approximation is consistent with a multiple-timescale interpretation, where $U^T$ evolves on a slow timescale while $V^T$ provides rapidly varying forcing.
This reduction provides a consistent link between the transient transverse dynamics and the steady kink-type structures analyzed below.
\subsection{Analytical solution for $U^T(y)$}
We consider the coupled system obtained by prescribing the spatial structure of the transverse velocity based on the dominant mode identified in Sec.~\ref{sec:transition}. 
Averaging over the fast temporal oscillations and retaining the dominant spatial structure, the transverse velocity is approximated by its leading-order spatial component:

\begin{equation} \label{eq:VTy}
V^T(y) = -\frac{\sin(n\pi y)}{n\pi \delta^2 Re},
\end{equation}
where $n$ is an integer, $Re$ is the Reynolds number, and $\delta$ is the characteristic length scale.

Introducing $W=dU^T/dy$, Eq.~(\ref{eq:UTy}) reduces to
\begin{equation}
\frac{dW}{dy} = Re\,V^T(y)\,W,
\end{equation}
with solution
\begin{equation}
W(y) = C_1 \exp\left(\frac{\cos(n\pi y)}{n^2\pi^2\delta^2}\right).
\end{equation}

Integration yields
\begin{equation}
U^T(y) = C_1 \int_0^y \exp\left(\frac{\cos(n\pi \xi)}{n^2\pi^2\delta^2}\right)\, d\xi + C_2.
\label{eq:UT_sol}
\end{equation}

The constants are determined from boundary conditions or normalization requirements depending on the physical interpretation of $U^T$.

\subsection{Asymptotic structure and kink-type solutions}

In the limit $\delta \ll 1$, define
\[
A_n = \frac{1}{n^2\pi^2\delta^2} \gg 1.
\]

The integral in Eq.~(\ref{eq:UT_sol}) is evaluated using Laplace's method. The dominant contributions arise from neighborhoods of the maxima of $\cos(n\pi y)$, located at
\[
y_k = \frac{2k}{n}.
\]

Expanding near $y_k$ gives
\[
\cos(n\pi y) \approx 1 - \frac{(n\pi)^2}{2}(y-y_k)^2,
\]
and therefore
\[
\exp(A_n \cos(n\pi y)) \approx \exp(A_n)\,
\exp\!\left(-\frac{(y-y_k)^2}{2\delta^2}\right).
\]

Substitution into Eq.~(\ref{eq:UT_sol}) yields the asymptotic form
\begin{equation}
U^T(y) \sim \sum_k C\, \mathrm{erf}\!\left(\frac{y-y_k}{\sqrt{2}\delta}\right),
\end{equation}
where $C \sim \delta\,\exp(A_n)$.

This result shows that the solution consists of a sequence of localized monotonic transitions of thickness $O(\delta)$, separating regions of nearly uniform velocity. Such structures are characteristic of kink-type solutions.

The spacing between adjacent structures is
\begin{equation}
\Delta y = \frac{2}{n}.
\end{equation}

\subsection{Connection to streak spacing}

Expressed in wall units,
\begin{equation}
\Delta y^+ = Re_\tau \frac{2}{n}.
\end{equation}

Using the experimentally observed value $\Delta y^+ \approx 100$ gives
\begin{equation}
n \approx \frac{2Re_\tau}{100}.
\end{equation}

For the parameters of Wei and Willmarth \cite{Wei_1989}, $\delta^+ \approx 10$ and $\delta=0.04$ yield $Re_\tau \approx 250$ and therefore $n \approx 5$, consistent with observed streak spacing.

\subsection{Streak intensity}

Near each transition point,
\begin{equation}
U^T(y) \sim C\, \mathrm{erf}\!\left(\frac{y-y_k}{\sqrt{2}\delta}\right),
\end{equation}
with amplitude $C \sim \delta\,\exp(A_n)$.

The corresponding velocity variation across the transition is
\begin{equation}
\Delta U^T \sim 2\delta\,\exp(A_n).
\end{equation}

In wall units,
\begin{equation}
\Delta U^{T+} \sim \frac{2\delta^+}{Re_\tau}
\exp\!\left(\frac{Re_\tau^2}{n^2\pi^2(\delta^+)^2}\right).
\end{equation}

The exponential factor reflects the sensitivity of the solution to $\delta$ and $n$, while the prefactor and global constraints limit the effective amplitude. As a result, the predicted streak intensities remain of order unity in wall units, consistent with experimental observations.

\subsection{ Streamwise extent of streaks}

The transient solution for the transverse velocity obtained in Sec.~\ref{sec:transition} exhibits oscillatory behavior with characteristic frequency $\lambda_n/2$. The corresponding temporal period is
\begin{equation}
\Delta T = \frac{2\pi}{\lambda_n}.
\end{equation}

Since the formation of streak-like structures is associated with the presence of transverse velocity fluctuations, the temporal scale $\Delta T$ provides an estimate of the characteristic lifetime of these structures. Assuming that the structures are convected downstream with a characteristic velocity $U_c$, their streamwise extent (streak length) can be estimated as
\begin{equation}
L_S \sim U_c \, \Delta T.
\end{equation}

Taking $U_c$ to be of the order of the local mean velocity (or friction velocity in wall units), this relation provides an estimate of the streak length in terms of the parameters of the model.
In wall units, this estimate becomes $L_S^+ \sim U_c^+ \Delta T^+$, which can be compared with experimentally observed streak lengths of order $10^3$ wall units.

Using the expression for $\lambda_n$, the characteristic time scale scales as $\Delta T \sim \delta Re$, yielding
\begin{equation}
L_S \sim U_c \delta Re.
\end{equation}
This scaling indicates that the streak length increases with Reynolds number, consistent with experimental observations.

\subsection{Remarks on boundary conditions}

The solution (\ref{eq:UT_sol}) is constructed to satisfy the governing equation (\ref{eq:UTy}). Imposing additional global constraints (such as enforcing zero mean over the domain) modifies the solution and may alter its local structure. In the present formulation, the focus is on the local behavior of $U^T(y)$, which governs the formation of kink-type transitions.

%
\section{Comparison with experimental observations}
\label{sec:discussion}

\subsection{Streamwise velocity: comparison with experiments}

Detailed comparisons between the analytical solution and experimental data have been reported in \cite{Fedoseyev_2025,Fedoseyev_2026}. Here we summarize the key results relevant to the present study.

\subsubsection{Wei and Willmarth channel flow experiments}

The experiments of Wei and Willmarth (1989) \cite{Wei_1989} were conducted in turbulent channel flow over the range $2970 \le Re \le 39582$ using distilled water.

\begin{figure}
\includegraphics[width=0.49\textwidth]{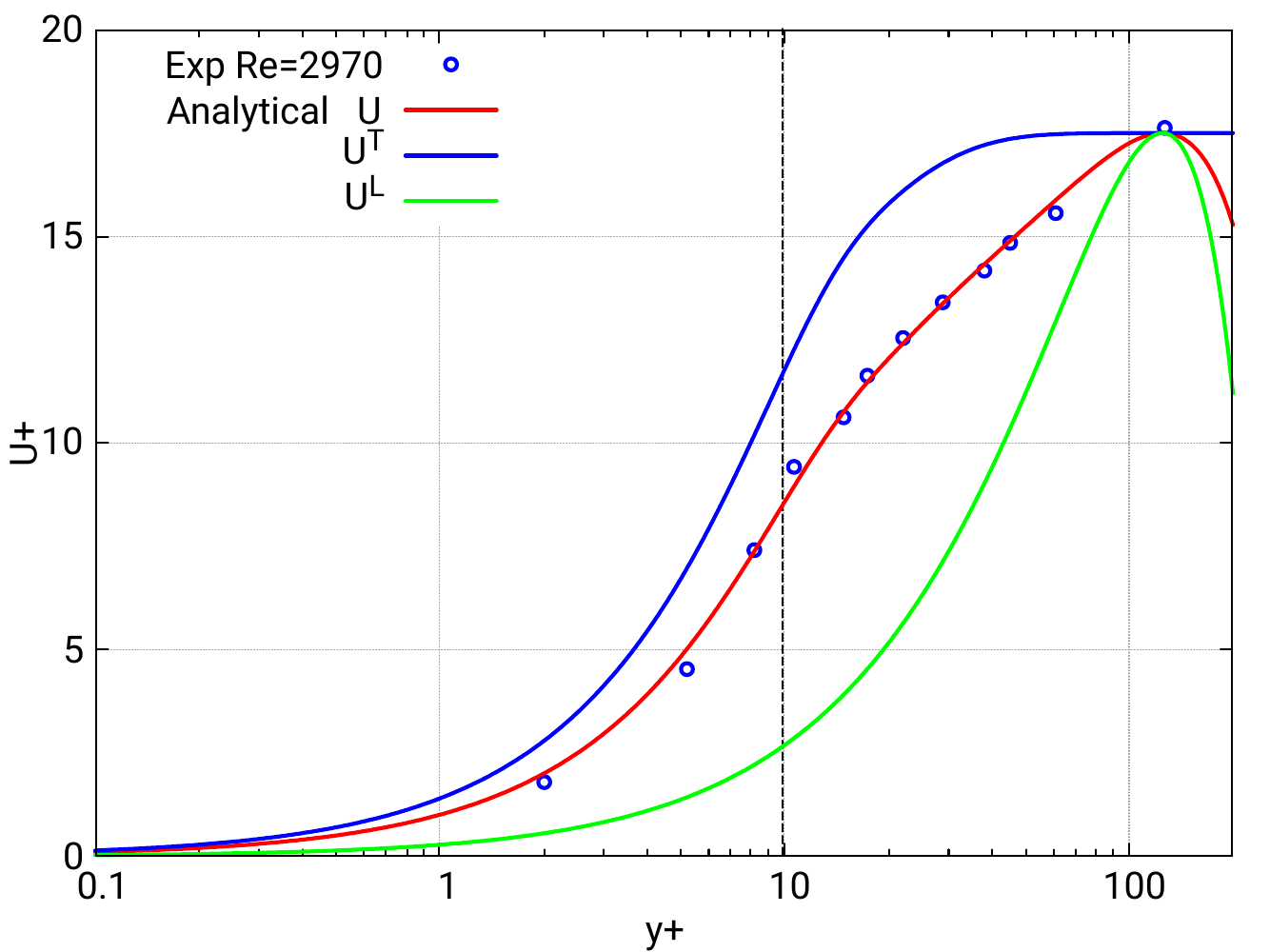}
\includegraphics[width=0.49\textwidth]{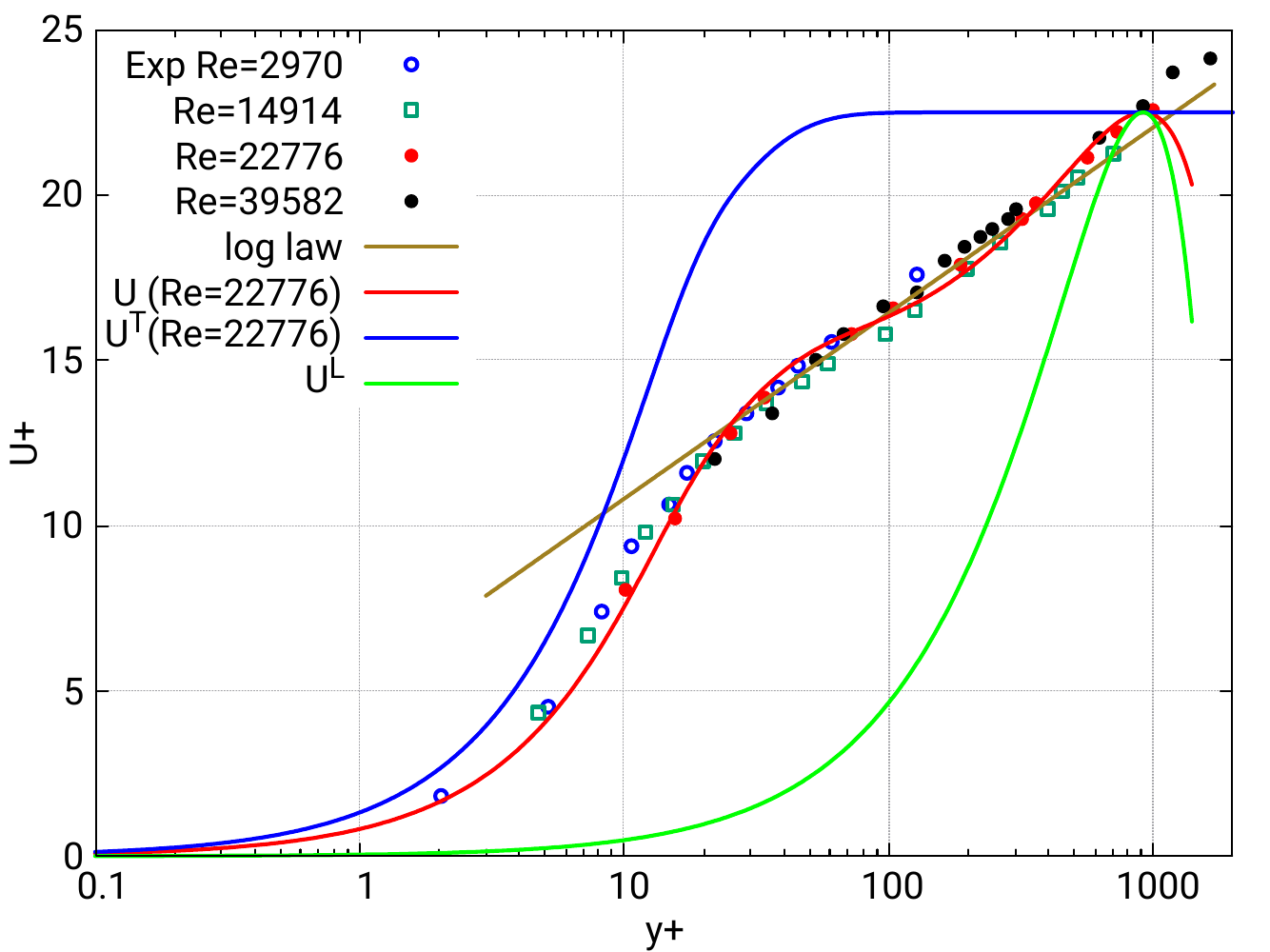}
\hspace{3cm} (a)    \hspace{8cm}(b)
\caption{\label{fig:comp_wei}(a)Comparison of streamwise mean velocity from  Wei (1989) experiments in $(y^+, U^+)$ coordinates at Re = 2970 (circles) with the laminar solution $U^L$ (green line),  the turbulent solution $U^T$ (blue line), 
and their superposition $U$ (solution of the Alexeev hydrodynamic equations, red line). The vertical dashed black line indicates $\delta$ at $y^+=10$.
(b) Streamwise velocity profiles for turbulent channel flow. Experimental data from Wei and Willmarth (1989) \cite{Wei_1989} are shown by symbols. The laminar solution $U^L$ (parabolic profile), the turbulent component $U^T$ (super-exponential form), and the combined AHE solution $U$  at $Re=22776$ are shown by green, blue, and red lines, respectively. The logarithmic law of the wall,  $U^+ = (1/\kappa)\log(y^+) + B$, is shown for reference (olive-green line). 
}
\end{figure}

Figure~\ref{fig:comp_wei}(a) shows the comparison between experimental data and the analytical solution at $Re=2970$ in wall coordinates $(y^+,U^+)$. The laminar component $U^L$ (parabolic profile), the turbulent component $U^T$, and the combined solution $U$ are shown. The parameter values $\gamma=0.65$ and $\delta=0.04$ are used. The vertical dashed line indicates $\delta^+ \approx 10$.

Figure~\ref{fig:comp_wei}(b) shows the comparison at $Re=22776$, including the logarithmic law of the wall for reference. The maximum deviation between the analytical solution and experimental data is within approximately $1\%$.
\begin{figure} 
\includegraphics[width=0.49\textwidth]{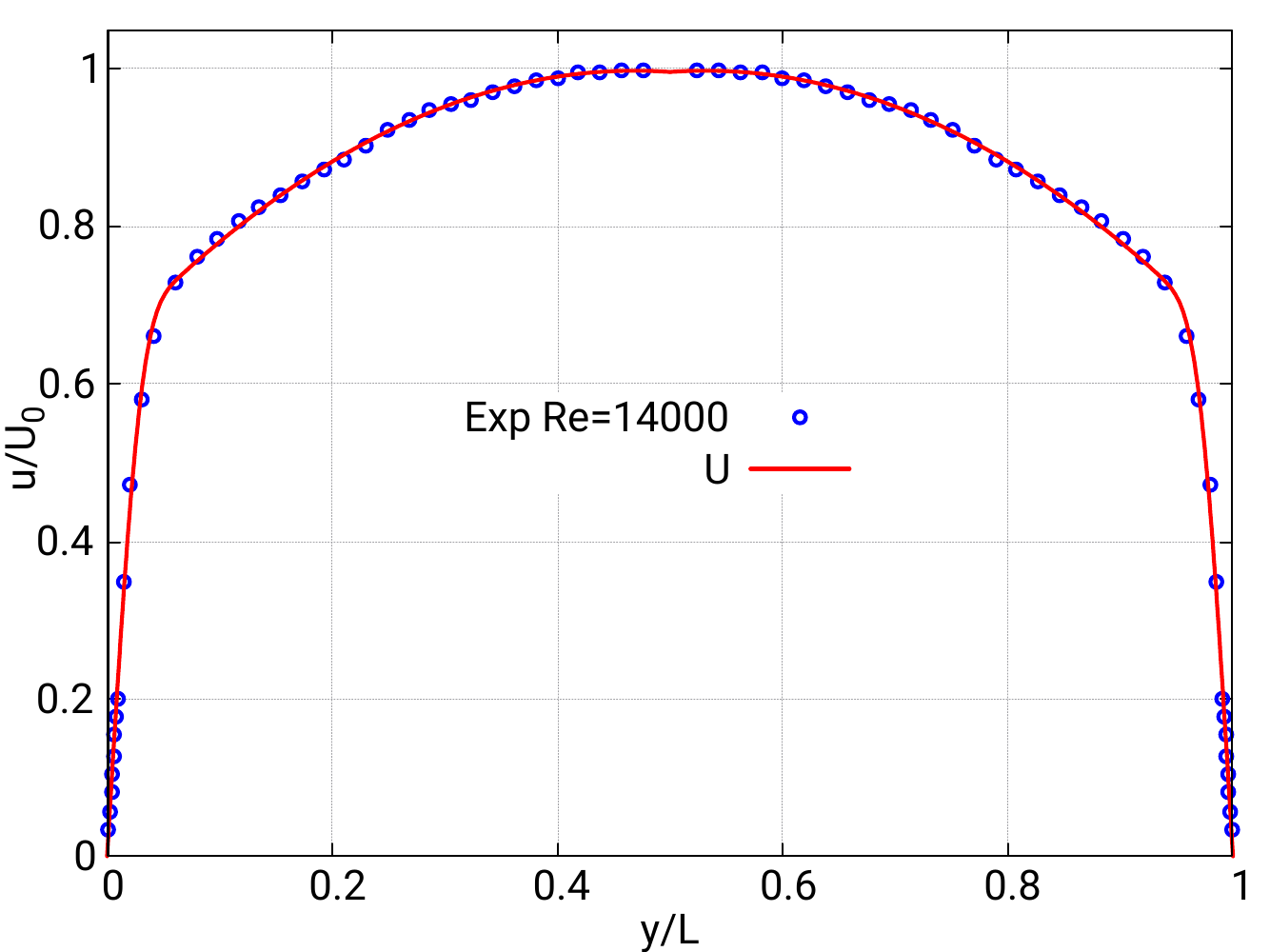}
\includegraphics[width=0.49\textwidth]{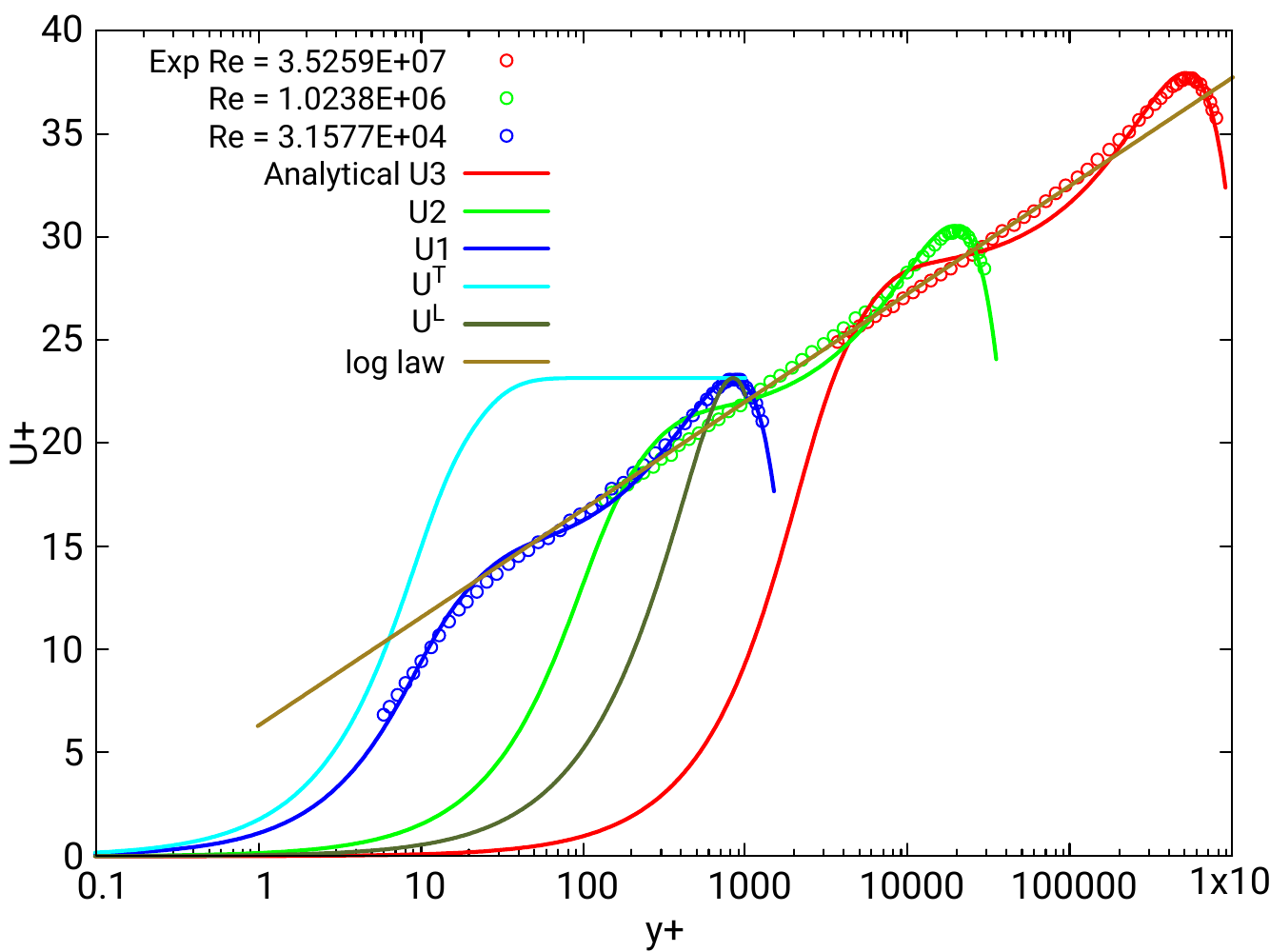}
\hspace{3cm} (a)    \hspace{8cm}(b)
\caption{
(a)  Comparison of streamwise velocity in turbulent channel experiment \cite{Pasch_2024}, $Re = 14000$, (blue circles) and analytical solution $U$ with coefficient $\gamma=0.65$ (red line).
(b) Comparison with turbulent pipe flow data from the Princeton Superpipe 
experiment \cite{Zagarola_1996} at $Re=3.15\times10^4$, $1.02\times 10^6$, and $3.52\times 10^7$ (symbols). 
The corresponding analytical solutions $U$ are shown by solid lines. The laminar ($U^L$) and turbulent ($U^T$) components are shown for $Re=3.15\times 10^4$ only. The logarithmic law is included for reference. The maximum error is approximately 3\% at the highest Reynolds number.}
\label{fig:pasch_zagarola}
\end{figure}

\subsubsection{Channel flow experiment by Pasch}

A turbulent channel flow experiment reported in \cite{Pasch_2024} was conducted at $Re=14000$ using air. The analytical solution with $\gamma=0.65$ and $\delta=0.027$ is compared with the measured velocity profile in Fig.~\ref{fig:pasch_zagarola}(a).

The agreement between the analytical solution and experimental data is within experimental uncertainty across the wall-normal range.

\subsubsection{Princeton Superpipe experiments}

A series of experiments was conducted in the Princeton Superpipe facility \cite{Zagarola_1996} over the range $3.15\times10^4 \le Re \le 3.52\times10^7$. 

Figure~\ref{fig:pasch_zagarola}(b) shows comparisons at three representative Reynolds numbers. The analytical solution captures the mean velocity profiles across a wide range of Reynolds numbers. The maximum deviation is approximately $3\%$ at the highest Reynolds number.

At very high Reynolds numbers ($Re > 10^6$), the Superpipe data may be influenced by surface roughness effects \cite{Barenblatt_1997}, which could contribute to the observed increase in deviation.

\subsection{Secondary flow: comparison with experiments}

Secondary flows in turbulent channels have been extensively studied \cite{Bradshaw_1987}. Measurements of transverse velocity in open-channel flows have been reported in \cite{Nezu_1984,Wang_2005,Wang_2006,Yang_2012,Chat_2023}.

Experimental observations indicate that the transverse velocity can be approximated by a sinusoidal profile of the form
\[
V \approx (0.01 - 0.02)U_0 \sin(2\pi y),
\]
which is consistent with the analytical form
\[
V^T = -\frac{\sin(2\pi y)}{2\pi \delta^2 Re}.
\]

\begin{figure}
\includegraphics[width=0.49\textwidth]{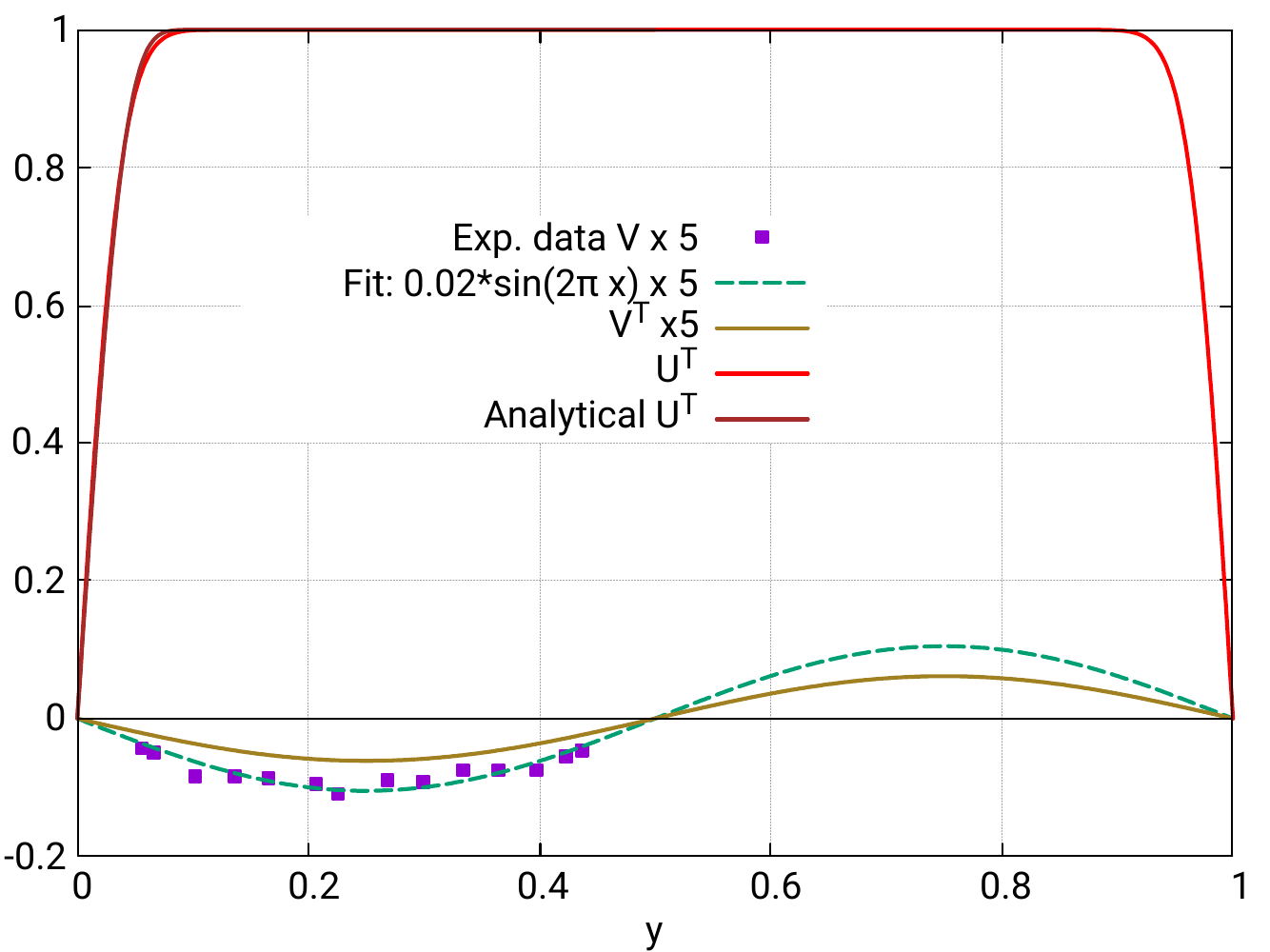}
\includegraphics[width=0.49\textwidth]{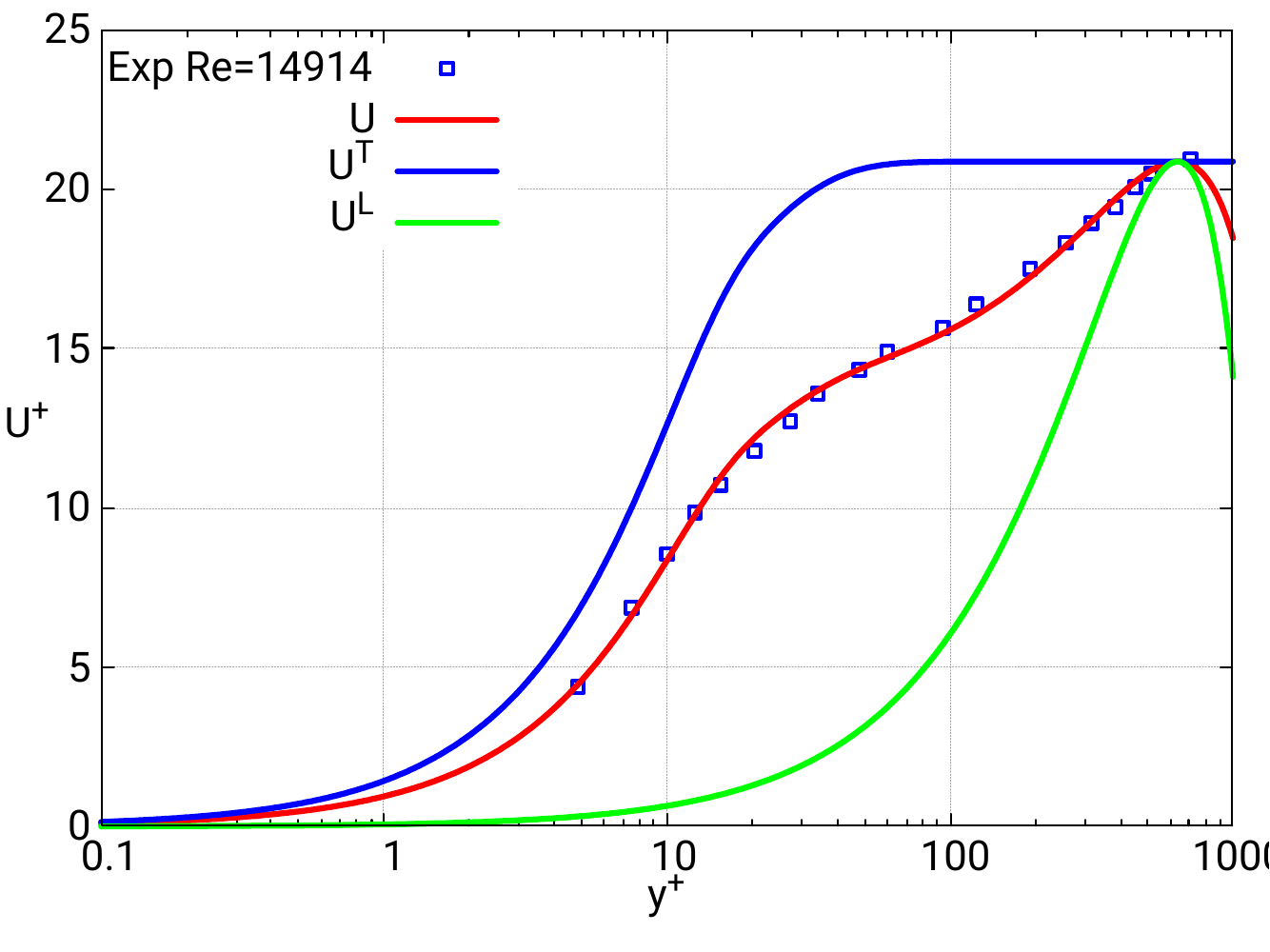}
\hspace{3cm} (a)    \hspace{8cm}(b)
\caption{\label{fig:vert}
(a) Comparison of vertical velocity data $V$ from \cite{Wang_2005} experiment (points), their fit by sinusoide (dashed green), and vertical velocity $V^T$ from Eq.(\ref{eq:VTy}) (olive-green). Also shown the corresponding streamwise velocity $U^T$, computed numerically and analytically from Eq.(\ref{eq:UTsol}) (red and brown)
(b) Comparison of streamwise mean velocity data from Wei (1989) experiments in $(y^+, U^+)$ coordinates at $Re = 14914$ (squares) with the laminar solution $U^L$ (green line), the turbulent solution $U^T$ is from Figure \ref{fig:vert}a (blue line), and their superposition $U$ (red line); $\delta=0.04$, $\gamma=0.65$.
}
\end{figure}

Figure~\ref{fig:vert}(a) shows a comparison between experimental data from \cite{Wang_2005} and the analytical profile. The agreement in both shape and amplitude supports the validity of the transverse velocity model.

The corresponding streamwise component $U^T$, computed from Eq.~(\ref{eq:UTsol}), is shown in Fig.~\ref{fig:vert}(b), along with experimental data from Wei (1989). The combined solution $U$ reproduces the measured profile with deviations within approximately $1\%$.

\subsection{Kink-type solutions}

Numerical solutions of Eq.~(\ref{eq:UTy}) were obtained for different values of the parameter $n$ in Eq.(\ref{eq:VTy}) using a finite-difference discretization. The resulting nonlinear system was solved using the Powell hybrid method \cite{Powell_1964}, with deflation techniques \cite{Farrell_2015} applied to avoid the trivial solution \cite{Fedoseyev_2026}.

\begin{figure}
\includegraphics[width=0.24\textwidth]{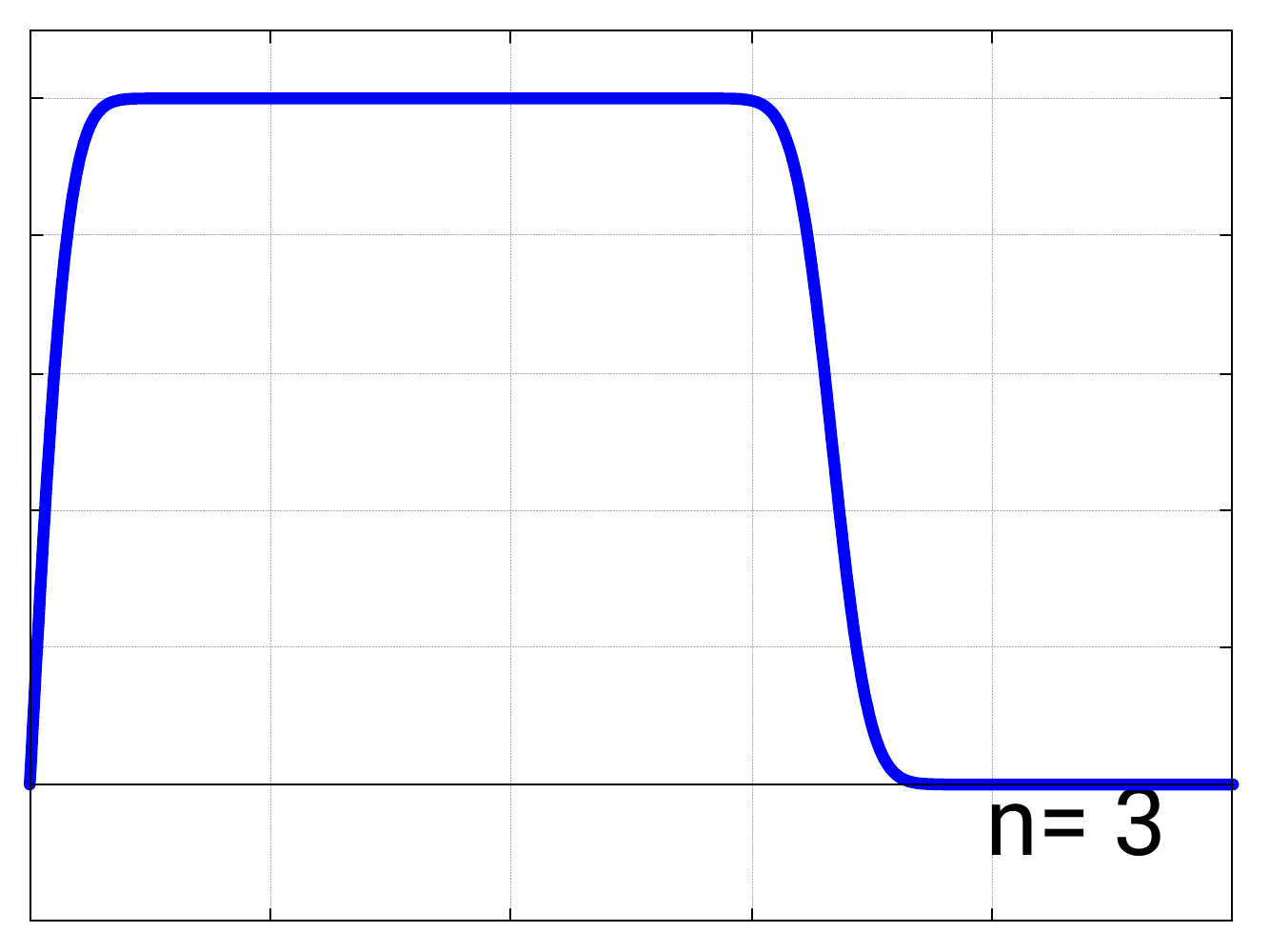}
\includegraphics[width=0.24\textwidth]{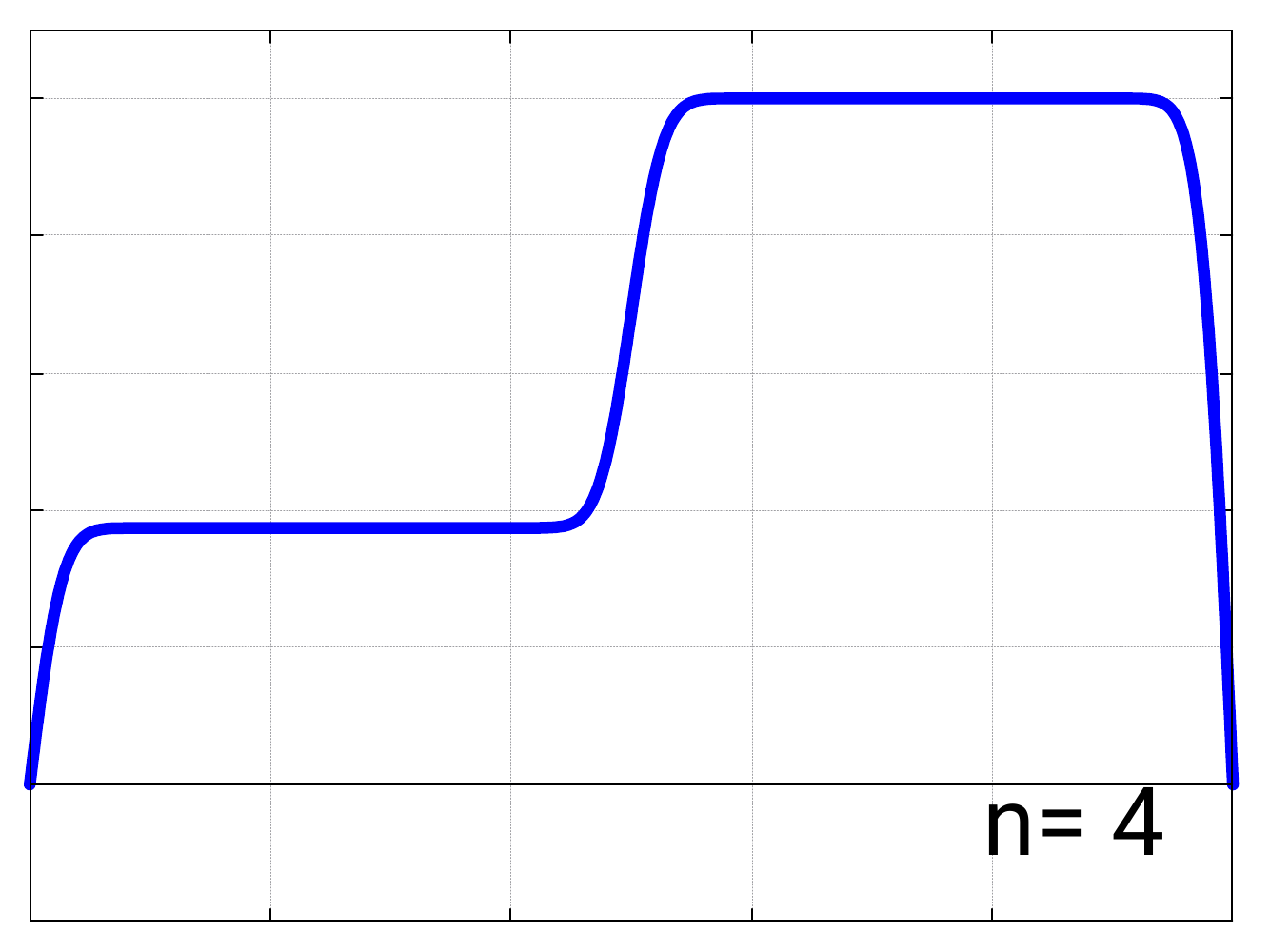}
\includegraphics[width=0.24\textwidth]{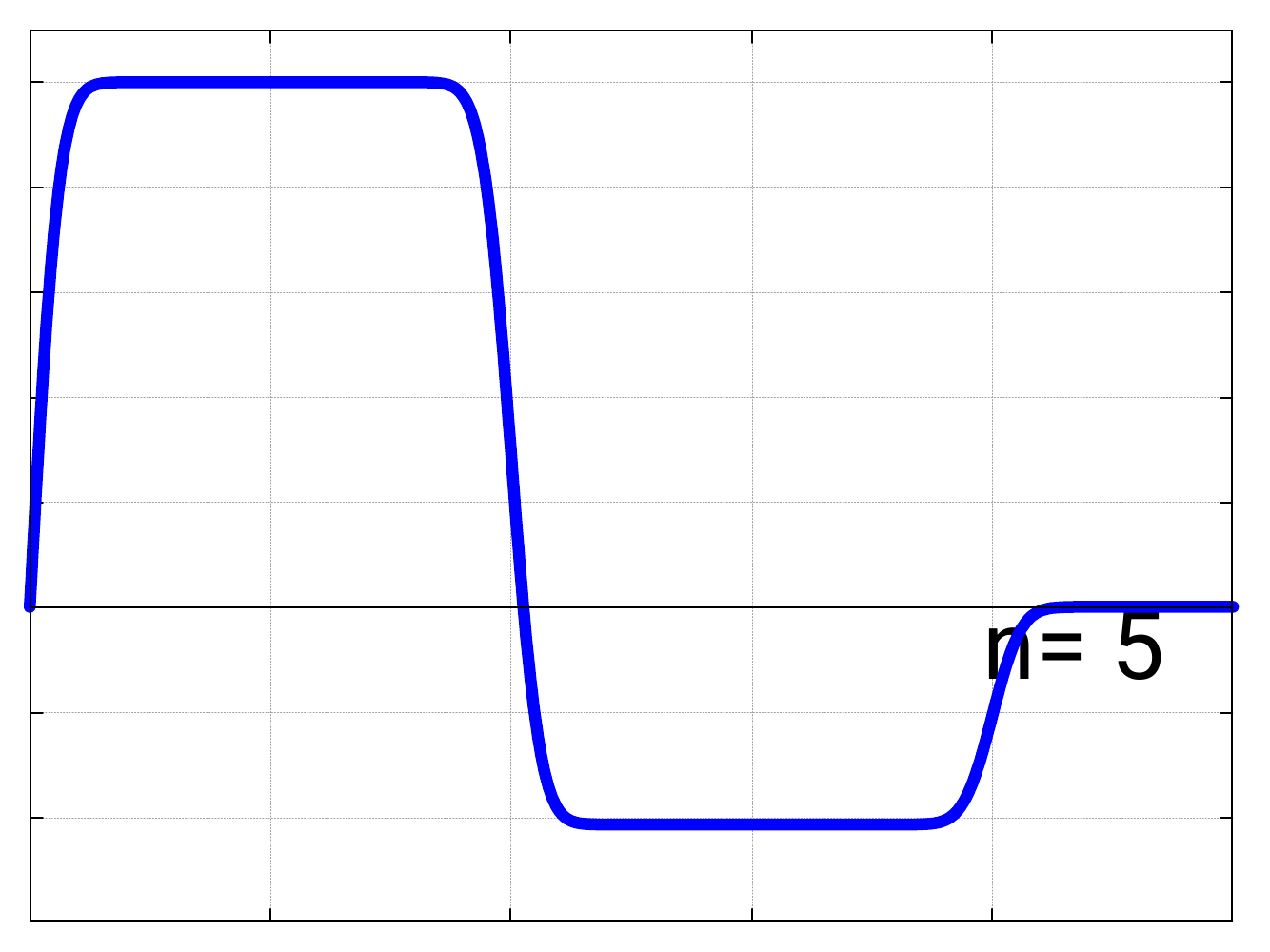}
\includegraphics[width=0.24\textwidth]{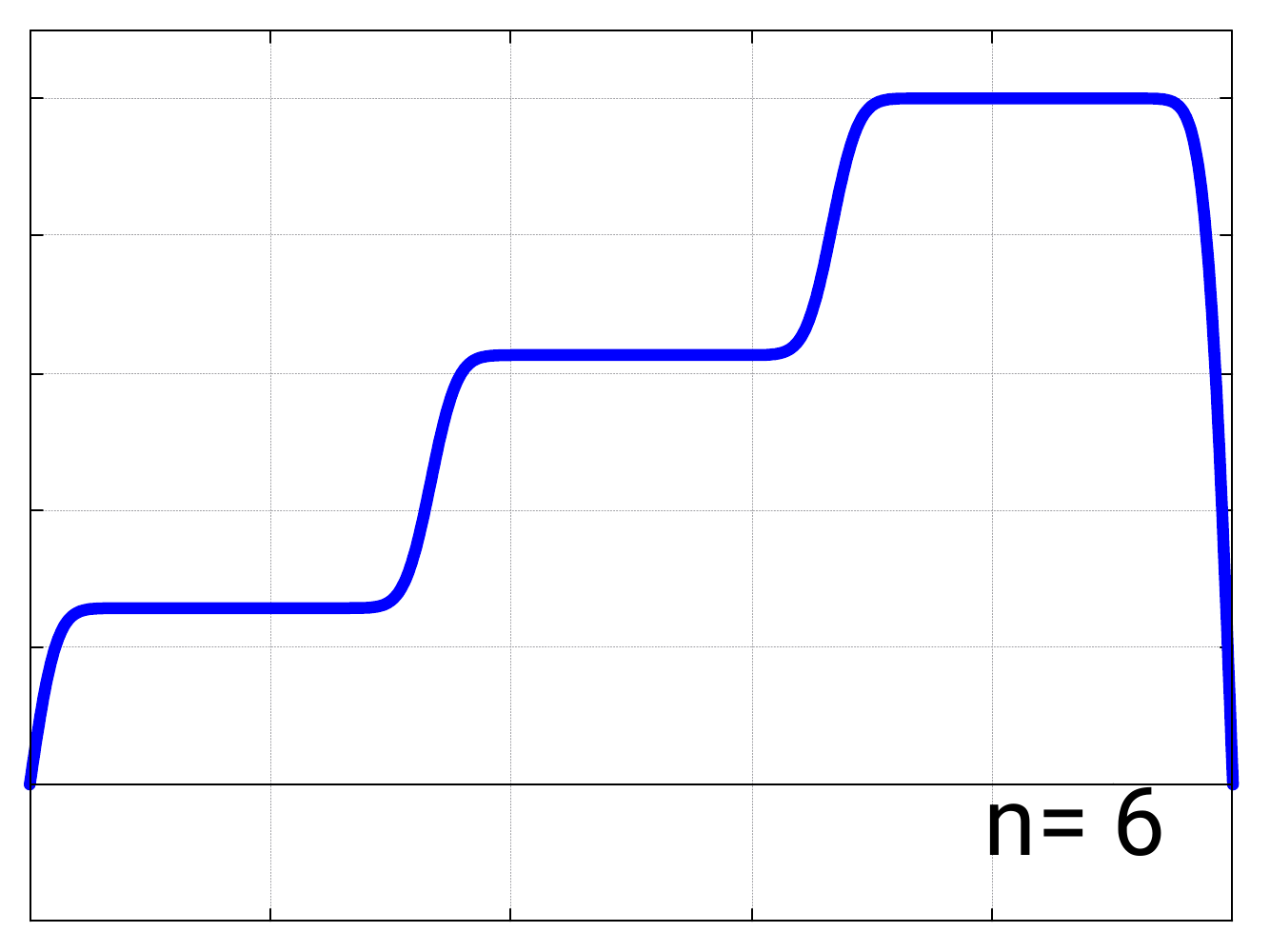}
\includegraphics[width=0.24\textwidth]{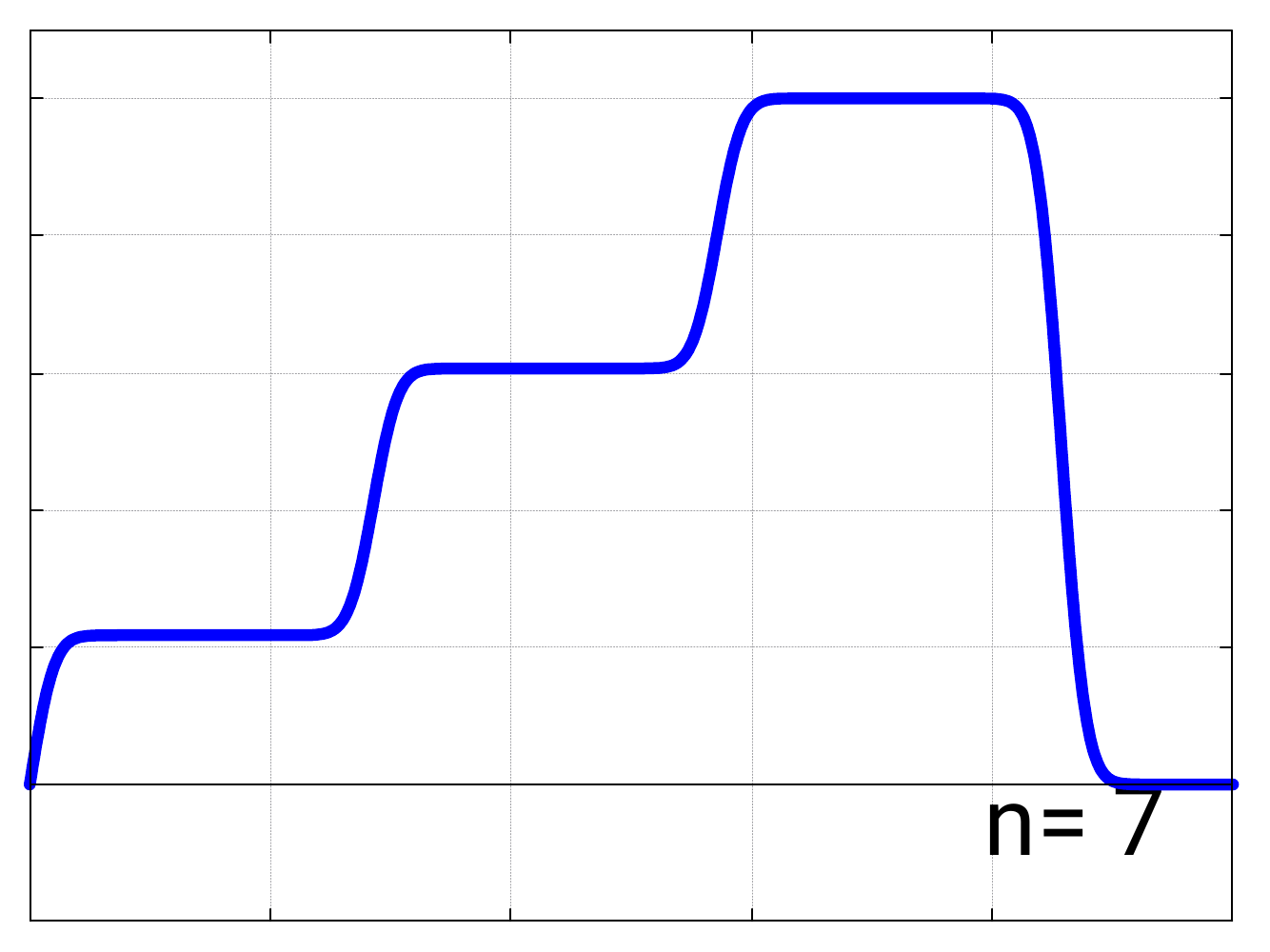}
\includegraphics[width=0.24\textwidth]{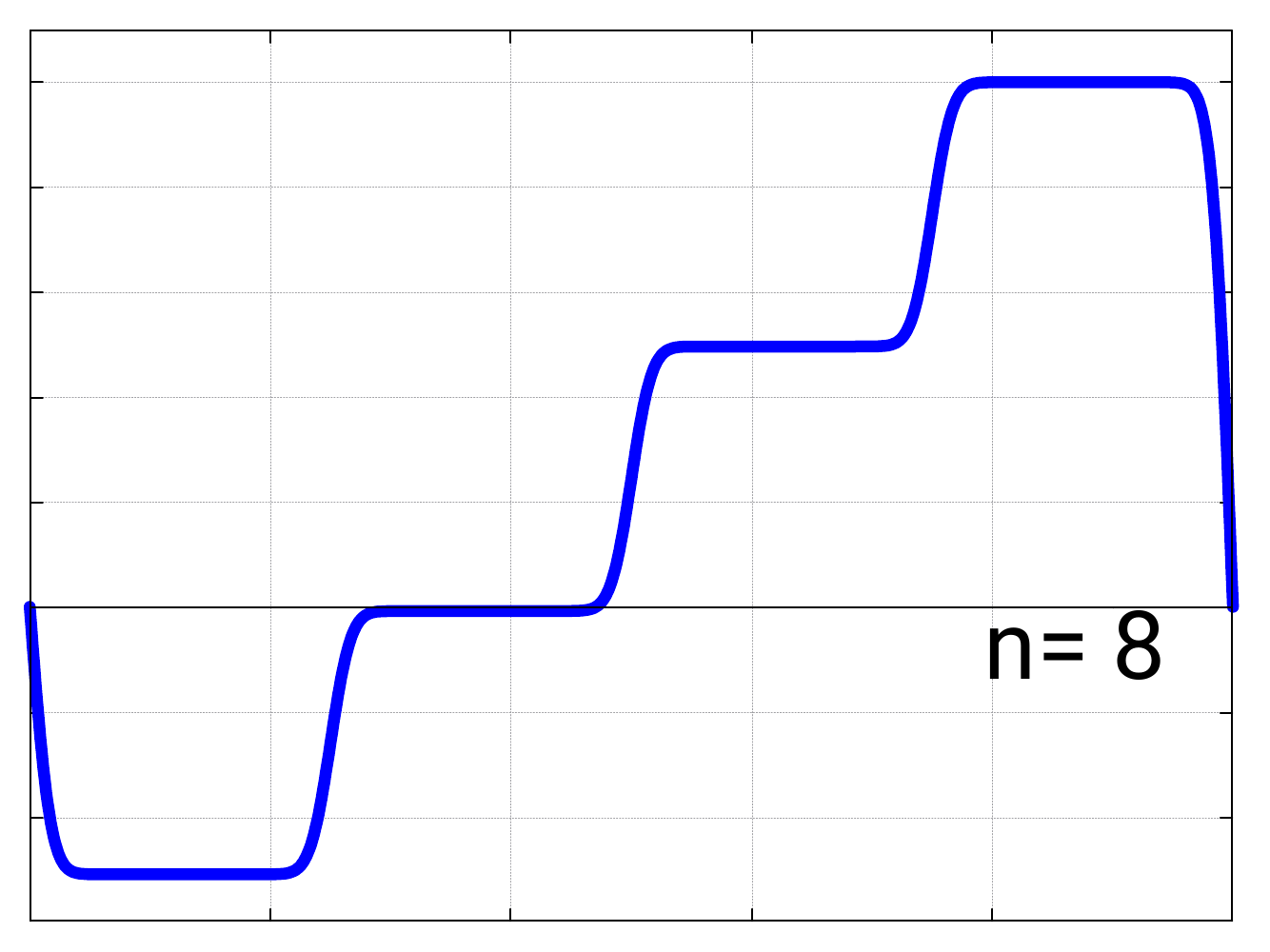}
\includegraphics[width=0.24\textwidth]{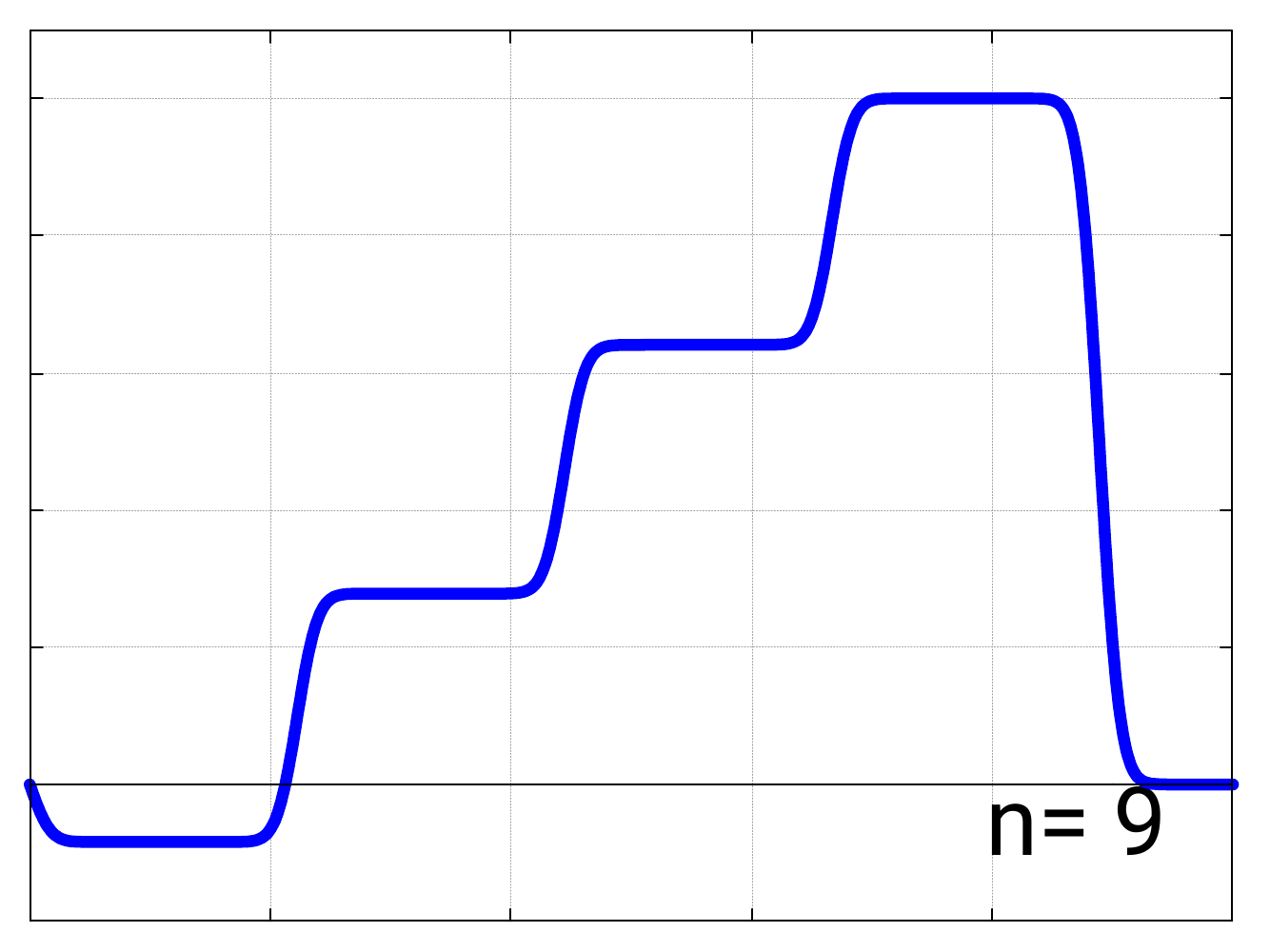}
\includegraphics[width=0.24\textwidth]{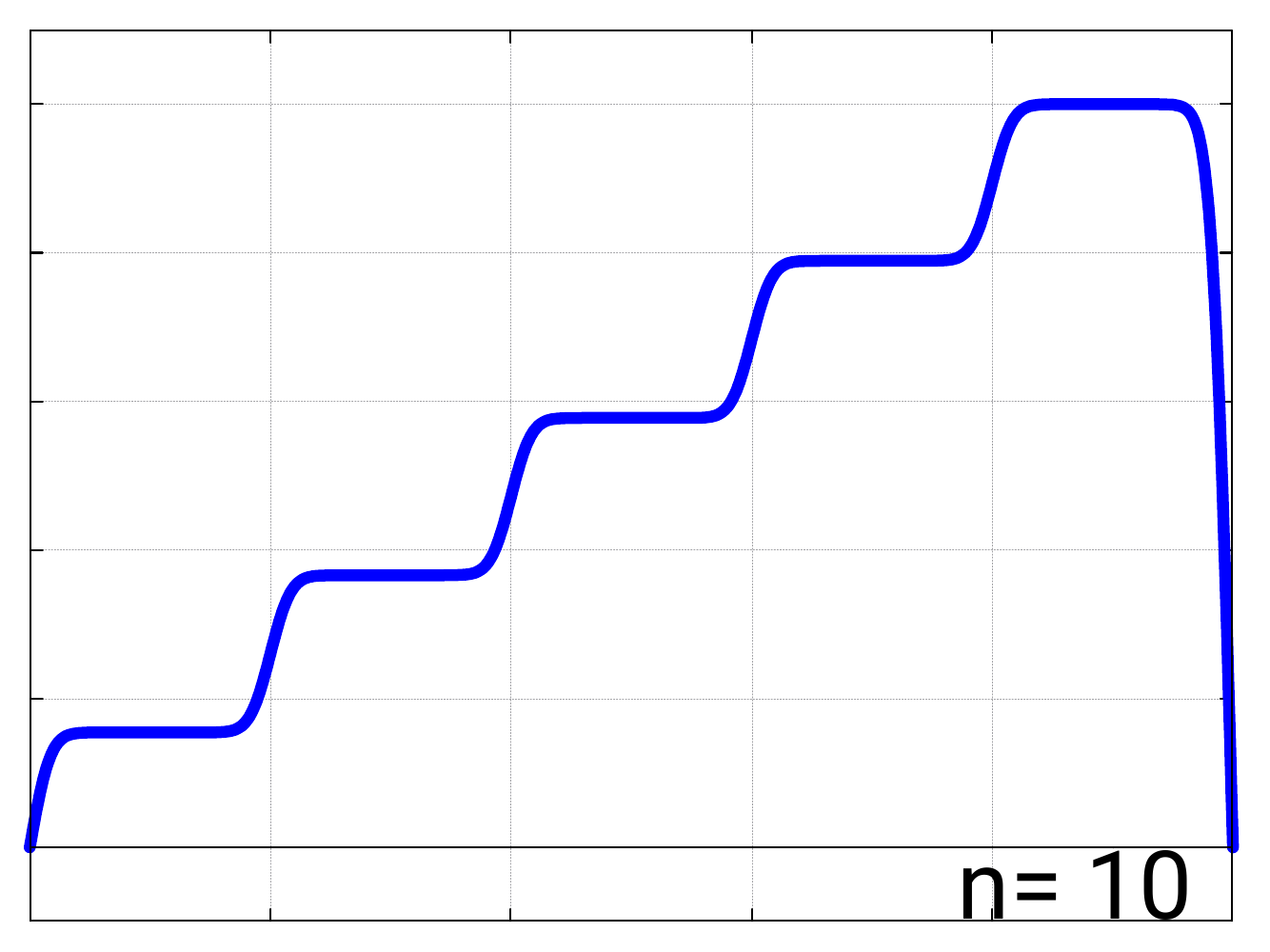}
\includegraphics[width=0.24\textwidth]{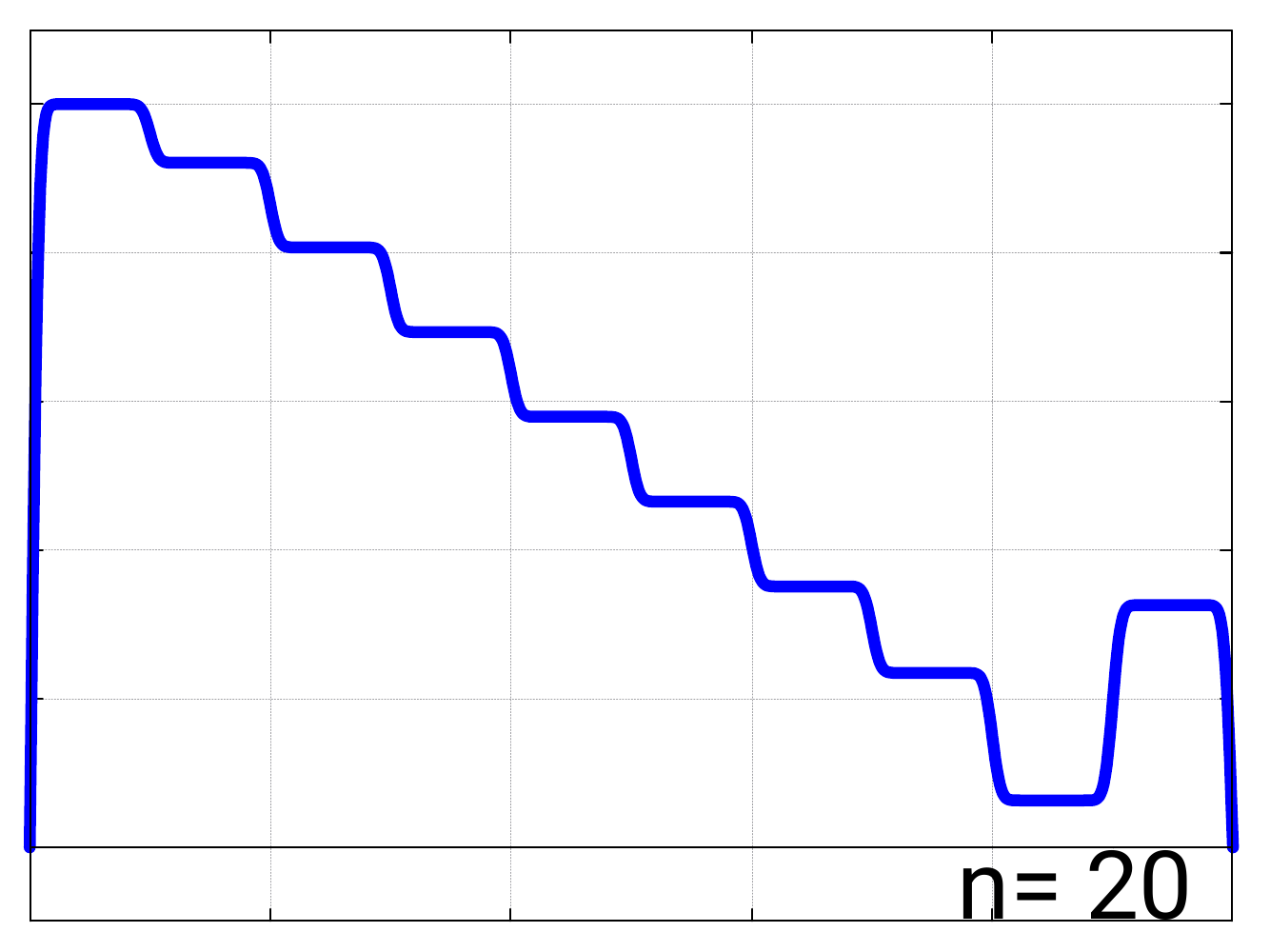}
\includegraphics[width=0.24\textwidth]{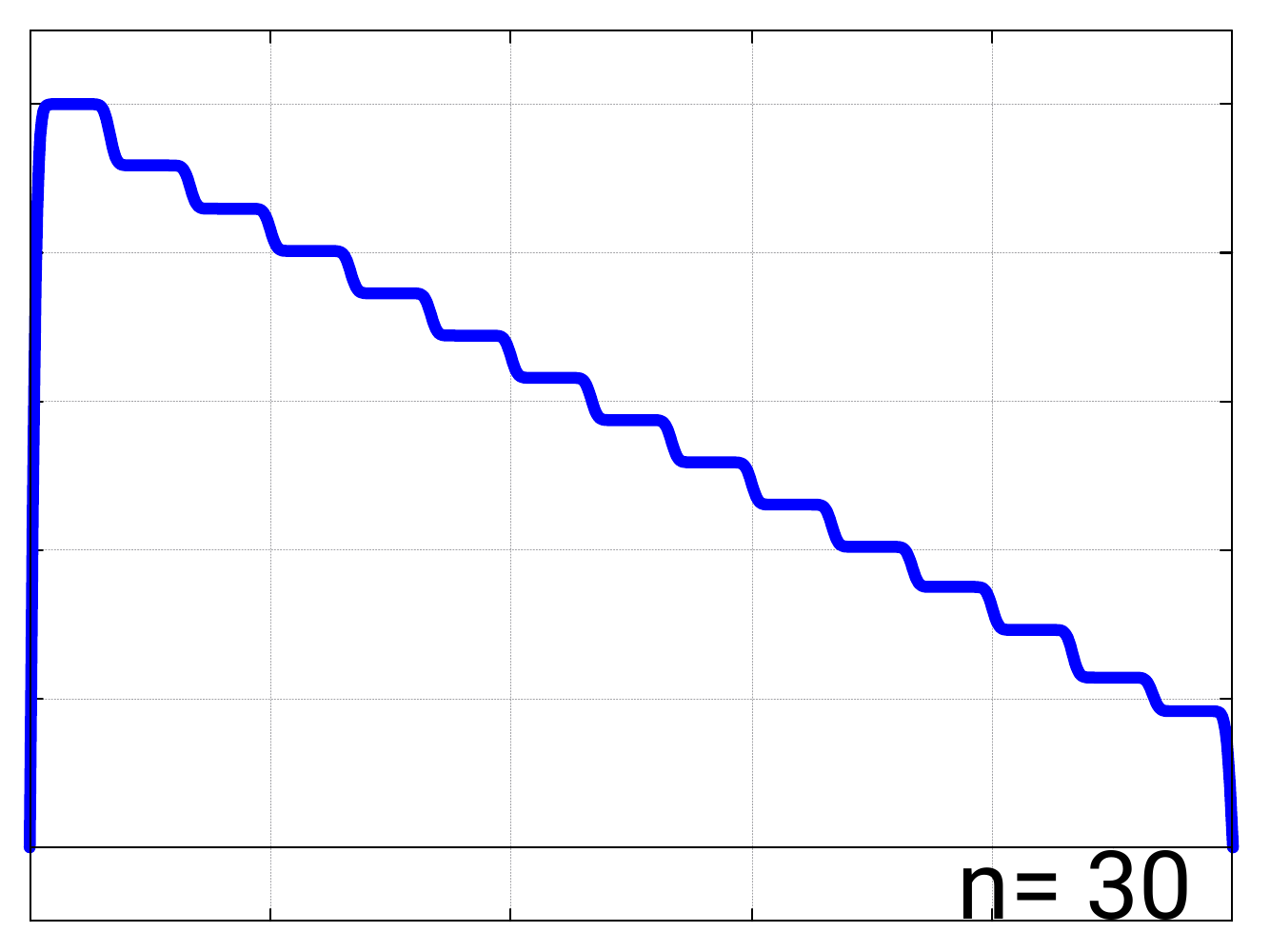}
\includegraphics[width=0.24\textwidth]{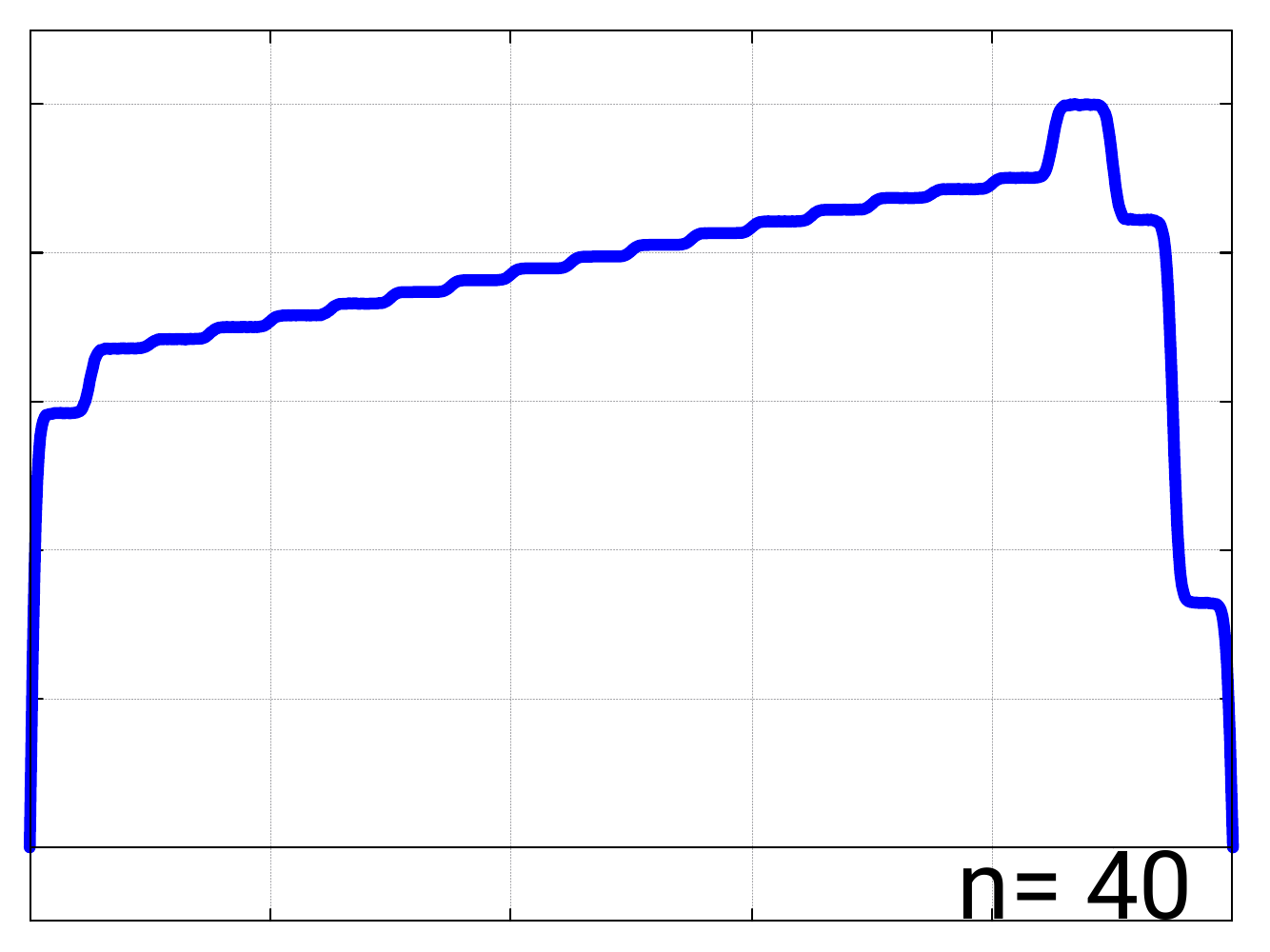}
\includegraphics[width=0.24\textwidth]{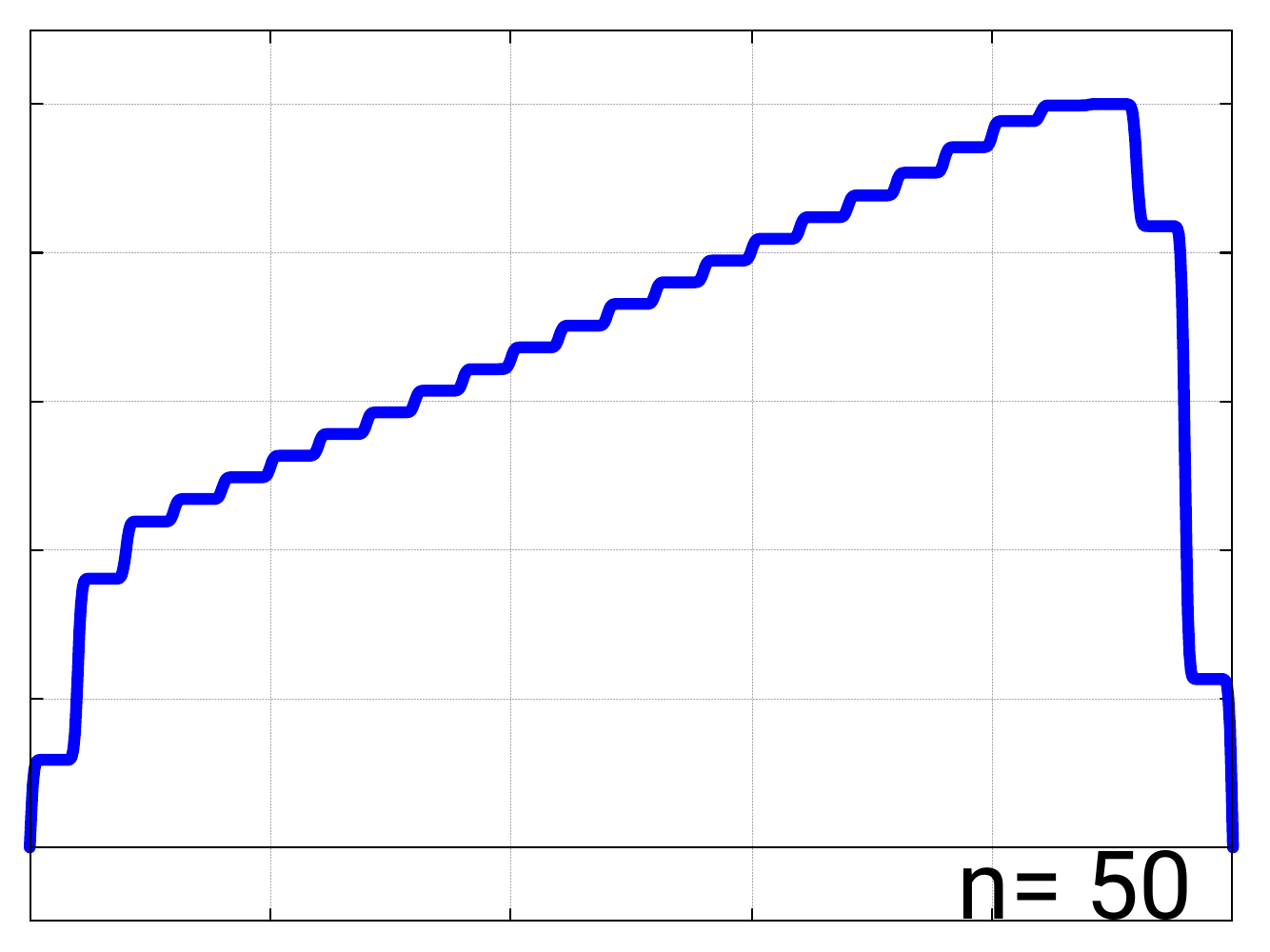}
\caption{\label{fig:kinks}
Kink-type solutions for $U^T$, Eq. (\ref{eq:UTy}), for different parameter $n$ in Eq.(\ref{eq:VTy}); $Re=14914$ and $\delta=0.04$. The verical axis is $U^T$, which is scaled to unit, and the horisontal axis is $y=[0,1]$.
}
\end{figure}

The computed solutions, shown in Fig.~\ref{fig:kinks}, confirm the analytical predictions of Sec.~\ref{sec:kink}. In particular, the solutions exhibit localized monotonic transitions of thickness $O(\delta)$ separating regions of nearly uniform velocity. The spacing between these transitions is consistent with the predicted value $\Delta y = 2/n$.

\subsection{Streak parameters}

An estimate of the streamwise extent of streak-like structures can be obtained from the temporal scale of the transverse velocity fluctuations. Using $\Delta T = 2\pi/\lambda_n$ and $L_S \sim U_c \Delta T$, a numerical estimate can be derived for the parameters corresponding to Wei and Willmarth \cite{Wei_1989}, with $Re=3000$, $\delta=0.04$, and $n\approx 5$.

This yields $\lambda_n \approx 0.14$ and $\Delta T \approx 44$. Taking a convection velocity $U_c \approx 0.04$ (corresponding to $U_c \sim 10u_\tau$ for $Re_\tau \approx 250$), the resulting streak length is $L_S \approx 1.8$, or $L_S^+ \approx 400$--$500$.

This estimate is consistent in order of magnitude with experimentally observed streak lengths, typically $O(10^3)$ wall units. The remaining discrepancy may be attributed to the simplified linear treatment of the transverse dynamics and the approximate choice of convection velocity.
%
%
\section{Conclusions}

An analytical framework for turbulent channel flow based on the Alexeev hydrodynamic equations has been developed and applied to the coupled dynamics of streamwise and transverse velocity components. The approach provides a unified description of mean velocity profiles, transverse velocity structure, and the emergence of streak-like features in wall-bounded turbulence.

An explicit analytical solution for the streamwise velocity was obtained as a superposition of laminar and turbulent components. The resulting velocity profiles show good agreement with experimental data from channel and pipe flows over a wide range of Reynolds numbers, with deviations typically within $1\%$ at moderate Reynolds numbers and up to approximately $3\%$ at the highest Reynolds numbers considered.

The transverse velocity component was analyzed using a linearized form of the governing equations. The resulting solution reveals oscillatory behavior with well-defined spatial modes. The dominant spatial structure is captured by a sinusoidal profile, which is consistent with experimentally observed secondary flows in turbulent channels and open flows.

A key result of the present work is the analytical characterization of the streamwise turbulent component as a class of kink-type solutions. In the asymptotic limit of small $\delta$, the solution exhibits localized monotonic transitions of thickness $O(\delta)$ separating regions of nearly uniform velocity. These structures form a periodic array with spacing $\Delta y = 2/n$, providing a direct analytical connection between the model parameter $n$ and experimentally observed streak spacing.

The model further predicts scaling laws for streak thickness, spacing, and intensity. In particular, the spacing expressed in wall units, $\Delta y^+ = 2Re_\tau/n$, yields values consistent with the experimentally observed spacing of approximately 100 wall units. The intensity of the structures is governed by an exponential sensitivity to $\delta$ and $n$, while remaining bounded due to global constraints.

An estimate of the streamwise extent of streaks was obtained from the characteristic temporal scale of transverse velocity fluctuations. The resulting streak length, expressed in wall units, is of order $10^2$--$10^3$, in agreement with experimental observations.

Overall, the present framework provides a consistent analytical description of several key features of wall-bounded turbulence, including mean velocity profiles, secondary flows, and streak formation. The results suggest that the coupling between transverse velocity and streamwise momentum plays a central role in the emergence of coherent structures in turbulent flows.

Future work will focus on incorporating nonlinear effects in the transverse dynamics and extending the analysis to more complex geometries and flow conditions.
%
%

%
\end{document}